\documentclass[conference]{IEEEtran}
\IEEEoverridecommandlockouts
\usepackage{array}
\usepackage[caption=false,font=normalsize,labelfont=sf,textfont=sf]{subfig}
\usepackage{textcomp}
\usepackage{stfloats}
\usepackage{url}
\usepackage{verbatim}
\usepackage{graphicx}
\usepackage{subcaption}
\usepackage{caption}
\usepackage{multirow}
\usepackage{pgfplots}
\usepackage{pgfplotstable}
\pgfplotsset{compat=1.16}
\usepackage[inkscapeformat=png]{svg}
\usepackage{adjustbox}
\usepackage{xcolor}

\usepackage{amsmath,amssymb,amsfonts}
\usepackage{textcomp}
\usepackage{xcolor}
\usepackage{enumitem}
\usepackage{booktabs}
\usepackage{colortbl}
\definecolor{lightgray}{gray}{0.9}
\usepackage{pifont}
\usepackage{algorithm}
\usepackage{algpseudocode}
\usepackage{makecell}
\usepackage{hyperref}
\usepackage{tabularx}
\newcolumntype{Y}{>{\centering\arraybackslash}X}

\makeatletter
\algrenewcommand\alglinenumber[0]{}
\makeatother

\def\BibTeX{{\rm B\kern-.05em{\sc i\kern-.025em b}\kern-.08em
    T\kern-.1667em\lower.7ex\hbox{E}\kern-.125emX}}

\begin{document}

\title{ROSA: Roundabout Optimized Speed Advisory with Multi-Agent Trajectory Prediction in Multimodal Traffic}

\author{\IEEEauthorblockN{Anna-Lena Schlamp\thanks{\copyright 2026 IEEE. Personal use of this material is permitted. Permission from IEEE must be obtained for all other uses, in any current or future media, including reprinting/republishing this material for advertising or promotional purposes, creating new collective works, for resale or redistribution to servers or lists, or reuse of any copyrighted component of this work in other works. This article has been accepted for publication in the proceedings of the \textit{2025 IEEE International Conference on Intelligent Transportation Systems (ITSC)}. This is the accepted manuscript version. The final version will be available in IEEE Xplore. $DOI:~10.1109/ITSC60802.2025.11423540$.}}
\IEEEauthorblockA{\textit{Institute AImotion Bavaria} \\
\textit{Technische Hochschule Ingolstadt}\\
Ingolstadt, Germany \\
anna-lena.schlamp@thi.de}
\and
\IEEEauthorblockN{Jeremias Gerner}
\IEEEauthorblockA{\textit{Institute AImotion Bavaria} \\
\textit{Technische Hochschule Ingolstadt}\\
Ingolstadt, Germany \\
jeremias.gerner@thi.de}
\and
\IEEEauthorblockN{Klaus Bogenberger}
\IEEEauthorblockA{\textit{School of Engineering and Design} \\
\textit{Technical University of Munich}\\
Munich, Germany \\
klaus.bogenberger@tum.de}
\and
\IEEEauthorblockN{Werner Huber}
\IEEEauthorblockA{\textit{CARISSMA Institute of Automated Driving
(C-IAD)} \\
\textit{Technische Hochschule Ingolstadt}\\
Ingolstadt, Germany \\
werner.huber@thi.de}
\and
\IEEEauthorblockN{Stefanie Schmidtner}
\IEEEauthorblockA{\textit{Institute AImotion Bavaria} \\
\textit{Technische Hochschule Ingolstadt}\\
Ingolstadt, Germany \\
stefanie.schmidtner@thi.de}
}

\maketitle

\begin{abstract}
We present ROSA -- Roundabout Optimized Speed Advisory -- a system that combines multi-agent trajectory prediction with coordinated speed guidance for multimodal, mixed traffic at roundabouts. Using a Transformer-based model, ROSA jointly predicts the future trajectories of vehicles and Vulnerable Road Users (VRUs) at roundabouts. 
Trained for single-step prediction and deployed autoregressively, it generates deterministic outputs, enabling actionable speed advisories. Incorporating motion dynamics, the model achieves high accuracy (ADE: 1.29m, FDE: 2.99m at a five-second prediction horizon), surpassing prior work. Adding route intention further improves performance (ADE: 1.10m, FDE: 2.36m), demonstrating the value of connected vehicle data. Based on predicted conflicts with VRUs and circulating vehicles, ROSA provides real-time, proactive speed advisories for approaching and entering the roundabout. Despite prediction uncertainty, ROSA significantly improves vehicle efficiency and safety, with positive effects even on perceived safety from a VRU perspective.
The source code of this work is available under: github.com/urbanAIthi/ROSA.
\end{abstract}


\section{Introduction}
Automated Driving (AD) promises safer, more efficient, and sustainable mobility. However, roundabouts remain a challenge to the AD stack due to dense, dynamic traffic and interaction-based driving behavior~\cite{Okumura2016ChallengesIP, PredictingBehaviourNaveed2019}. 
Complexity further intensifies in multimodal traffic with Vulnerable Road Users (VRUs, i.e., pedestrians and cyclists), whose behavior is highly variable and difficult to predict. Unlike human drivers, Automated Vehicles (AVs) cannot rely on eye contact or body language as a fallback to resolve uncertain situations, which poses a risk to the safe coexistence of AVs and VRUs at roundabouts~\cite{Rasouli2017AgreeingTC, Giuffr2012UnderstandingSI}. 

To improve VRU safety and predictability, infrastructure-based solutions such as prioritized zebra crossings or signalized crosswalks have been proposed~\cite{Meneguzzer2013AnalysisAC, XiaoZhaoLu2010MultimodalAO}. While effective in reducing risks, this static batching of road users causes wasteful stops and inefficient gap usage -- reducing roundabout capacity, increasing delays, and raising emissions~\cite{Meneguzzer2013AnalysisAC, CapaEffectsTollazzi2008}. 
Coordinating vehicle behavior in response to prioritized crossing VRUs can reduce inefficiencies while preserving safety. 
Proactive speed reduction has also been shown to build trust and improve perceived safety from a VRU perspective \cite{Dey2017TheIO, Schneemann2016AnalyzingDI}. However, such coordination depends on accurate prediction of VRU behavior.
This paper proposes ROSA, a Roundabout Optimized Speed Advisory system. ROSA integrates interaction-aware trajectory prediction into coordinated speed guidance for multimodal traffic. It is designed to support both automated and human-driven vehicles, aiming to safely and efficiently interact with VRUs at roundabouts.

\begin{table*}[tb] 
\centering
\captionsetup{justification=centering}
\caption{Comparison of related prediction models by design aspects and performance.}
\resizebox{\textwidth}{!}{
\begin{tabular}{|c|c|c|c|c|>{\centering\arraybackslash}p{1.0cm}|>{\centering\arraybackslash}p{1.2cm}|>{\centering\arraybackslash}p{1.0cm}|>{\centering\arraybackslash}p{1.2cm}|c|c|}
\hline
\multirow{2}{1.5cm}{\centering  \textbf{Paper}} & \multirow{2}{1.5cm}{\centering  \textbf{Model}} & \multirow{2}{1.5cm}{\centering  \textbf{Roundabout}} & \multirow{2}{1.0cm}{\centering  \textbf{Vehicles}} & \multirow{2}{0.7cm}{\centering  \textbf{VRUs}} &  \textbf{Multi-Agent} & \textbf{Agent Dynamics} & \textbf{History Length} & \textbf{Prediction Horizon} & \multirow{2}{1.5cm}{\centering  \textbf{ADE / FDE}} \\
\hline
 Vanilla-TF \cite{Quintanar2021PredictingVT} & TF & \checkmark & \checkmark & -- & -- & \checkmark & -- & 5.0 & 1.88 / 4.85 \\
 IAMP \cite{Trentin2023LearningenabledMM} & Dynamic Bayesian Network & \checkmark & \checkmark & -- & -- & \checkmark & 4.0 & 4.0 & 1.43 / 3.75 \\
 3D-A-W \cite{Hasan2021ManeuverbasedAT} & Anchor Trajectories + LSTM & \checkmark & \checkmark & -- & -- & \checkmark & 2.0 & 4.0 & -- / 3.08 \\
 GCN \cite{Zhang2024ExploringRD} & Gated Recurrent Unit & \checkmark & \checkmark & -- & -- & \checkmark & 1.0 & 1.0 & 0.16 / 0.31 \\
 ITRA \cite{Scibior2021ImaginingTR} & RNN & \checkmark & \checkmark & -- & \checkmark & \checkmark & 1.0 & 3.0 & 0.17 / 0.49 \\
 MTP-GO \cite{Westny2023MTPGOGP} & Temporal GNN & \checkmark & \checkmark & -- & \checkmark & \checkmark & 3.0 & 5.0 & 0.97 / 3.02 \\
 mmTransformer \cite{Liu2021MultimodalMP} & TF & -- & \checkmark & -- & -- & -- & 2.0 & 3.0 & 0.84 / 1.34 \\
  TF \cite{Giuliari2020TransformerNF} & TF & -- & -- & \checkmark & -- & -- & 3.2 & 4.8 & 0.31 / 0.55 \\
GSG-Former \cite{Luo2023GSGFormerGS} & Attention-based GNN + Temporal TF & -- & -- & \checkmark & -- & \checkmark & 3.2 & 4.8 & 10.27 / 16.67 \\
 AMENet \cite{Cheng2020AMENetAM} & Attentive Maps Encoder Network & -- & \checkmark & \checkmark & -- & \checkmark & 3.2 & 4.8 & 0.73 / 1.59 \\
 DCENet \cite{Cheng2020ExploringDC} & Dynamic Context Encoder Network & -- & \checkmark & \checkmark & -- & \checkmark & 3.2 & 4.8 & 0.52 / 1.23 \\
 SCOUT \cite{Carrasco2021SCOUTSA} & Attention-based GNN & \checkmark & \checkmark & \checkmark & \checkmark & -- & 3.2 & 4.8 & 1.38 / 3.45 \\
 GNN \cite{Daoud2023PredictionOA} & Attention-based GNN & \checkmark & \checkmark & \checkmark & \checkmark & -- & 3.0 & 3.0 & -- / 1.7 \\
\hline
\end{tabular}
}
\label{table:prediction_models}
\end{table*}

\subsection{Coordination at Roundabouts} \label{sec:coordination}
To increase efficiency and traffic flow at roundabouts, several works propose a coordination among vehicles in a fully automated setting. 
By means of different techniques, such as forming clusters \cite{BiLevelBakibillah2021} or determining sequence and speed trajectories~\cite{OptControlZhao2017, TrajektorienplanerLong2022}, the vehicles are steered in a way that their driving behavior, i.e., speed or acceleration, is optimized. 
All studies successfully reduce travel and waiting times, fuel consumption, and emissions, depending on traffic conditions and demand. 
However, existing approaches neglect VRUs in both the coordination logic and evaluation, limiting applicability in multimodal traffic. Even accelerating toward the roundabout is encouraged if a cluster or a gap can be caught, which may compromise perceived safety from a VRU perspective. 
Previous works assume cooperative driving behavior in a fully automated setting. Based on their premises, a full AV penetration rate is required to achieve improvements~\cite{OptControlZhao2017, Montanaro2018TowardsCA}, which restricts applicability in mixed-traffic environments.
Rather than predicting future traffic states, they analytically solve optimization problems based on the current situation, lacking an integrated prediction and coordination framework. Moreover, the works are not grounded in real-world data and do not quantify efficiency or safety gains in realistic traffic scenarios.

\subsection{Trajectory Prediction at Roundabouts}  
Proactive speed coordination in response to conflicting vehicles and VRUs requires accurate trajectory prediction to estimate their future positions. Several approaches exist, trained on real-world datasets and varying in methods and architectural design. Most adopt an ego-centric perspective \cite{Quintanar2021PredictingVT, Hasan2021ManeuverbasedAT, Liu2021MultimodalMP}, where an ego vehicle predicts future states of surrounding agents within its field of view. In contrast, so-called multi-agent approaches use a bird's-eye view perspective to jointly predict motion of all agents, capturing interdependencies and improving accuracy \cite{Scibior2021ImaginingTR, Westny2023MTPGOGP, Carrasco2021SCOUTSA, Daoud2023PredictionOA}.
Common model architectures include Graph Neural Networks (GNNs), Recurrent Neural Networks (RNNs), such as Long Short-Term Memory (LSTM), and Transformers (TFs). RNNs are limited by their sequential processing, while GNNs rely on prior assumptions (e.g., graph construction or map inputs). In contrast, Transformers offer a more flexible, data-driven approach with minimal structural constraints \cite{Vaswani2017AttentionIA}.
Many works produce multi-trajectory predictions to handle uncertainty in motion and intent, capturing a probability distribution over possible future trajectories rather than a single deterministic path \cite{Trentin2023LearningenabledMM, Westny2023MTPGOGP, Liu2021MultimodalMP, Luo2023GSGFormerGS, Cheng2020AMENetAM}. Prediction performance is typically assessed using Average Displacement Error (ADE) and Final Displacement Error (FDE), which quantify deviations from ground-truth trajectories over time.

Table \ref{table:prediction_models} compares related models by reported ADE/FDE (in meters) and key design aspects. 
Most studies use a three-second input history and predict five seconds ahead. Recent approaches increasingly incorporate heterogeneous agent dynamics due to their proven benefit for accuracy \cite{Westny2023MTPGOGP, Cheng2020ExploringDC}. 
However, multi-agent models remain underrepresented, despite their strength in modeling inter-agent dependencies. Moreover, the joint modeling of vehicles and VRUs in roundabout scenarios is particularly underexplored. Models like AMENet \cite{Cheng2020AMENetAM}, and DCENet \cite{Cheng2020ExploringDC} consider both agent types, but do not focus on roundabout scenarios in their evaluation. Models with more extensive roundabout evaluations often focus on a single agent type \cite{Quintanar2021PredictingVT, Zhang2024ExploringRD}, while combined models tend to perform worse at roundabouts than at intersections \cite{Carrasco2021SCOUTSA, Daoud2023PredictionOA}. This underlines the need for accurate, joint prediction of vehicles and VRUs at roundabouts. Multi-agent models and the consideration of heterogeneous agent dynamics are essential for capturing mutual interactions \cite{Westny2023MTPGOGP}. Additionally, route-level information could help reduce prediction uncertainty, but has not yet been explored in current models.

\subsection{Contributions}

We propose ROSA -- Roundabout Optimized Speed Advisory -- a system designed for coordinated speed guidance in multimodal traffic.
While current coordination approaches neglect VRUs, ROSA enables real-time, proactive vehicle speed adjustments in response to both prioritized crossing VRUs and intersecting vehicles when approaching and entering roundabouts. The goal is to reduce wasteful stops and delays, improving vehicle efficiency, while also enhancing overall traffic safety and perceived safety from a VRU perspective through minimized conflicts and proactive speed reduction. 
Unlike prior coordination methods focused on fully automated, cooperative driving, our approach is also designed for mixed traffic environments with automated and human-driven vehicles.
At its core, ROSA integrates interaction-aware trajectory prediction to anticipate the future traffic state and support informed speed advisories. By developing a multi-agent, VRU-aware prediction model that accounts for distinct agent dynamics and route information, this work addresses the gap of jointly predicting the trajectories of vehicles and VRUs at roundabouts, a setting underexplored in existing research. 
The main contributions of this work can be summarized as follows:
\begin{itemize}
\item[1.] An interaction-aware prediction model is developed that jointly predicts trajectories of vehicles and VRUs at roundabouts using real-world data. The model operates in a multi-agent and autoregressive manner to support integration with real-time speed optimization. 
\item[2.] The impact of including agent-specific dynamics and route information on prediction accuracy is evaluated. Moreover, the model’s ability to predict conflict zone occupancy -- essential for speed optimization -- is assessed.
\item[3.] The speed advisory of ROSA is introduced to proactively coordinate vehicle behavior in response to predicted conflicts with VRUs and other vehicles in multimodal and mixed traffic environments. Its effectiveness in enhancing vehicle efficiency and safety is evaluated through simulations based on real-world trajectory data.
\end{itemize}
\noindent
We consider a VRU-prioritized setting, which is common in urban safety-focused designs \cite{Meneguzzer2013AnalysisAC}. ROSA is trained and evaluated on real-world trajectory data. In practice, trajectories of approaching vehicles and VRUs, necessary to run ROSA, can be obtained via infrastructure sensors or sensors built into modern connected vehicles, acting as so-called Floating Car Observers (FCOs) \cite{gerner2023FCO} and transmitting their information via Vehicle-to-Everything (V2X) communication. To support further research, we make the source code of this work available under: github.com/urbanAIthi/ROSA.

\section{Interaction-Aware Multi-Agent Trajectory Prediction} \label{sec:prediction}
For the interaction-aware trajectory prediction of all agents within or approaching the roundabout (see Fig. \ref{fig:Kreisel}), we employ a Transformer architecture. It leverages self-attention to capture inter-agent dependencies and dynamics directly from raw trajectory data, without relying on prior assumptions (e.g., graph generation in GNNs, maps). The model is trained for single-step prediction using a three-second history, and deployed autoregressively to generate multi-step trajectories over a prediction horizon of five seconds. Unlike multi-trajectory prediction methods, we generate single, deterministic outputs to provide actionable speed advisories in ROSA.

\subsection{Dataset and Preprocessing} \label{dataset}
\begin{figure}[tb]
    \centering
    \includegraphics[width=0.9\columnwidth]{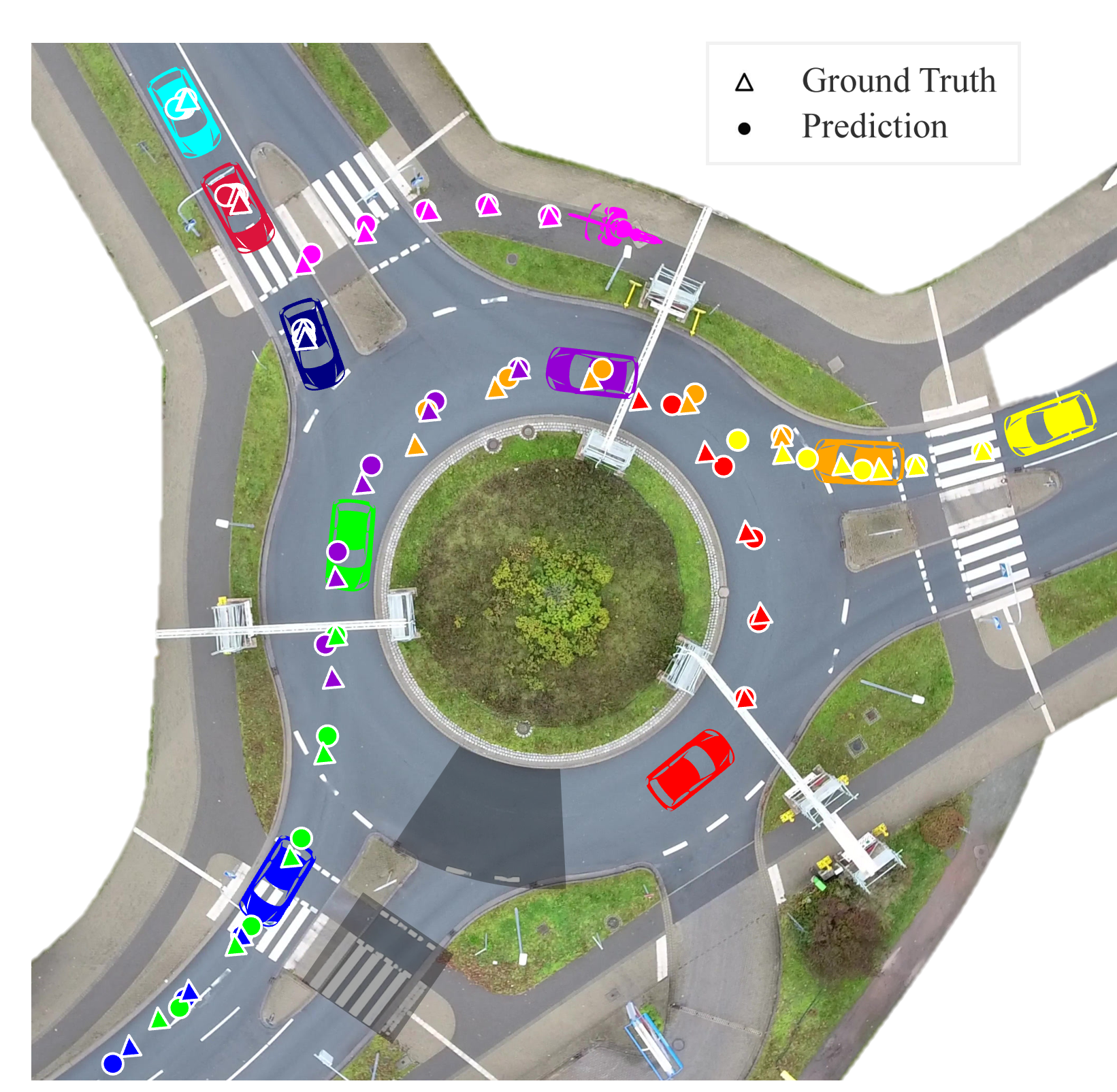}
    \caption{Visualization of the urban roundabout \textit{rdb1} from the \textit{openDD} dataset \cite{Breuer2020openDDAL} with prioritized VRU crossings. The sample scenario illustrates interaction-aware trajectory prediction by ROSA jointly for all agents, including a cyclist (magenta). Shaded areas highlight a representative crosswalk and entry conflict zone for occupancy evaluation.}
    \label{fig:Kreisel}
\end{figure}

Trajectory prediction at roundabouts represents a significant challenge due to their complex and inherently multimodal interactions, particularly involving VRUs. Several datasets have been introduced to address this challenge, including \textit{rounD}~\cite{Krajewski2020TheRD}, \textit{INTERACTION} \cite{Zhan2019INTERACTIONDA}, and \textit{openDD}~\cite{Breuer2020openDDAL}. In this work, we particularly focus on the \textit{openDD} dataset, which provides 30~Hz drone-recorded trajectories of naturalistic roundabout traffic with kinematic and semantic annotations. We hereby focus on \textit{rdb1} (see Fig. \ref{fig:Kreisel}), selected for its prioritized VRU crossings at all entries and exits, enabling frequent and consistent multimodal interactions. We extend the original dataset with route information (potentially accessible through V2X-enabled navigation data), inferring each agent's exit intention based on its final position relative to the roundabout center. Agents remaining inside the roundabout or classified as pedestrians/cyclists are labeled $-1$, others receive an exit arm label. To prepare the model input, irrelevant features are removed, retaining only the class label $c$ indicating vehicle or VRU, the positional coordinates $p_x$ and $p_y$, the velocity $v$, the tangential and lateral accelerations \( a_{\text{tan}} \) and \( a_{\text{lat}} \), the heading angle $\theta$ and the exit $e$. 
To reduce temporal redundancy and prepare for second-wise prediction, the dataset is downsampled to 1~Hz, with agent states aggregated per second. The dataset is subsequently divided into 80\% training, 10\% validation, and 10\% test sets, ensuring a balanced representation of VRUs across all splits.


\subsection{Model Architecture} \label{sec:model}
To predict the future positions of all road users at the roundabout, we utilize the Transformer-based architecture proposed in \cite{gerner2025FCO_TFCO}. In contrast to conventional trajectory prediction algorithms, which typically employ an ego-centric perspective, this approach captures a holistic view of the traffic environment. Consequently, it is particularly effective in modeling the complex interactions among road users within the unique structural dynamics of a roundabout. The architecture is founded on embedding all available kinematic and semantic information of agents within a specified region of the traffic network, denoted as $N$. Initially, this data is normalized with respect to the considered traffic segment and subsequently transformed into an embedding space using a multilayer perceptron~(MLP). Agent information from the preceding $s$ time steps is also embedded to effectively capture the dynamic behavior of the system. Positional embeddings are added to encode temporal information, explicitly indicating each embedding's corresponding time step, ranging from $0$~(current) to $-s$ (past). Thus, the input comprises a sequence of $N~\times~s$ embeddings processed by a Transformer Encoder architecture. An attention mask is applied to reduce the quadratic computational complexity inherent to the attention mechanism. This mask restricts each embedding of agent $N_i$ at time step $t$ to attend only to embeddings of the same agent across all time steps and embeddings of all agents at the same time step. Consequently, the number of attention connections is reduced from $(N~\times~s)^2$ to $N~\times~s+s$, while still effectively capturing both the dynamics of individual agents and their interactions with other agents. The network output consists of the processed embeddings from the masked Transformer Encoder at the current time step. These embeddings are subsequently passed through an MLP prediction head to yield predicted kinematic properties. In contrast to the original architecture, which predicts only positions ($\hat{p_x}$, $\hat{p_y}$), our approach extends the prediction to include additional kinematic attributes for each agent. Specifically, the model predicts velocity ($\hat{v}$), longitudinal acceleration ($\hat{a}_{\text{tan}}$), lateral acceleration ($\hat{a}_{\text{lat}}$), and orientation represented through the sine and cosine of the heading angle ($\sin(\hat{\theta})$, $\cos(\hat{\theta})$). Consequently, for each agent $N_i$ at the current timestep, the model outputs a tuple:
\[
\hat{\mathbf{r}} = 
\left[ 
\hat{p}_x,\ \hat{p}_y,\ 
\hat{v},\ 
\hat{a}_{\text{tan}},\ \hat{a}_{\text{lat}},\ 
\sin(\hat{\theta}),\ \cos(\hat{\theta})
\right].
\]
The model predictions are compared against ground-truth data, and deviations are quantified through a composite loss function. The position and speed predictions are evaluated using Mean Squared Error (MSE), accelerations are assessed with Smooth L1 loss to increase robustness against outliers, and orientation predictions utilize MSE along with a geometric regularization term ensuring unit-length orientation vectors. The final loss per agent aggregates these individual components using weighted summation, emphasizing accurate position prediction while incorporating auxiliary motion dynamics.

\subsection{Autoregressive Prediction}
The Transformer-based prediction model described above is originally designed for single-step prediction, forecasting the immediate future kinematic properties of all agents at the roundabout. However, for applications like speed advisory systems for approaching vehicles, knowledge about the traffic evolution over a longer time horizon is essential to enable anticipatory and efficient behavior planning. To address this requirement, we extend the single-step prediction architecture into an autoregressive framework. In this setup, the model remains unchanged during training and continues to predict the agent states at $t+1$ based on historical input sequences. During inference, the predicted kinematic states are recursively fed back as inputs for subsequent prediction steps, thereby generating multi-step trajectories into the future.
More formally, the model predicts all agent states at time step $t+1$ based on an input sequence spanning time steps $[t-s,\, t]$ of variable length $s \in [s,\, s+m-1]$, where $m$ denotes the prediction horizon. At each autoregressive step, predictions are compared to ground truth for loss computation and evaluation, and the input sequence is updated by appending the latest prediction. This procedure allows us to forecast the roundabout traffic state over extended horizons while leveraging the same model architecture and training process.

\subsection{Evaluation Setup}

Following prior work in the field, we evaluate the Transformer-based model on the test set using ADE and FDE.
To systematically evaluate the contribution of different input modalities, we conduct an ablation study comparing a baseline model that leverages only positional and type data against variants progressively enriched with motion dynamics and exit intention. Each configuration modifies the input feature vector~$i$ accordingly, with associated effects in the prediction target $\mathbf{r}$ and output $\hat{\mathbf{r}}$ (see Section~\ref{sec:model} for details). The tested input configurations are:

\begin{enumerate}[label=\arabic*.]
    \item \textbf{Position and Type:} \\
    $i = [c, p_x, p_y]$ 
    \item \textbf{+ Dynamics:} \\
    $i = [c, p_x, p_y, v, a_{\text{tan}}, a_{\text{lat}}, \sin(\theta), \cos(\theta)]$
    \item \textbf{+ Exit Intention:} \\
    $i = [c, p_x, p_y, v, a_{\text{tan}}, a_{\text{lat}}, \sin(\theta), \cos(\theta), e]$
\end{enumerate}
\smallskip

As an additional performance metric, related to the goal of coordinated speed guidance at roundabouts, we evaluate how well predicted agent positions translate into accurate predictions of crosswalk occupancy by VRUs and entry occupancy by circulating vehicles within the roundabout. Based on the roundabout's topology, we define three occupancy zones for crosswalks and three for entries (see Fig.~\ref{fig:Kreisel}, with one sample crosswalk and entry zone visualized), yielding 19,944 binary classifications per prediction step and zone type across the test dataset. To evaluate performance, we define an occupied zone as a positive case (P) and a free zone as a negative case (N), enabling the use of standard binary classification metrics to assess the quality of occupancy determination based on the predicted trajectories.

\subsection{Results}
The evaluation results in Table~\ref{table:results} show that the baseline model (1) exhibits substantial error growth over time, reaching an ADE of 5.08m and an FDE of 10.51m at a five-second prediction horizon. 
This indicates that positional and agent type information alone is not sufficient for reliable motion prediction in complex, interaction-dependent environments like roundabouts.
Augmenting the input with dynamic features~(2) significantly improves prediction accuracy across all time steps. At five seconds, ADE declines to 1.29m and FDE to 2.99m, already outperforming previous works on roundabout trajectory prediction for both vehicles and VRUs \cite{Carrasco2021SCOUTSA, Daoud2023PredictionOA} (see Table \ref{table:prediction_models}).
Incorporating exit intention (3) yields further improvements, especially for longer horizons, reducing ADE and FDE to 1.10m and 2.36m. These findings highlight the importance of motion dynamics and route awareness, supporting the integration of connected vehicle data where available.

\begin{table*}[tb] 
\centering
\captionsetup{justification=centering}
\caption{Comparison of trajectory prediction accuracy for different input feature configurations. Average Displacement Error (ADE) and Final Displacement Error (FDE) (in meters) are reported over prediction horizons ranging from~1~to~5~seconds.}
\begin{tabular}{|c|c|c|c|c|c|}
\hline 
\multirow{2}{*}{\textbf{Input Features}} 
& \multicolumn{5}{c|}{\textbf{ADE / FDE}} 
\\ 
& \textbf{1} & \textbf{2} & \textbf{3} & \textbf{4} & \textbf{5}  
\\
\hline  
\textbf{Position and Type (1)} 
& 0.70 / 0.70 & 1.54 / 2.37 & 2.54 / 4.55 & 3.73 / 7.28 & 5.08 / 10.51
\\
\textbf{+ Dynamics (2)} 
& 0.14 / 0.14 &  0.30 / 0.46 & 0.54 / 1.01 & 0.86 / 1.84 & 1.29 / 2.99
\\
\textbf{+ Exit Intention (3)}
& 0.14 / 0.14 & 0.30 / 0.46 & 0.52 / 0.96 & 0.79 / 1.59 & 1.10 / 2.36
\\ 
\hline 
\end{tabular}
\label{table:results}
\end{table*}

\renewcommand{\arraystretch}{1.2}
\begin{table}[tb] 
\centering
\captionsetup{justification=centering}
\caption{Occupancy estimation performance for crosswalk and entry zones at prediction horizons ranging from 1 to 5~seconds. Results for model~(2) (motion dynamics) are shown in the top row, and for model~(3) (including exit intention) in the bottom row.}
\begin{tabular}{|>{\centering\arraybackslash}m{1.3cm}|
                >{\centering\arraybackslash}m{1.3cm}
                >{\centering\arraybackslash}m{1.3cm}|
                >{\centering\arraybackslash}m{1.3cm}
                >{\centering\arraybackslash}m{1.3cm}|}
\hline
\multirow{2}{*}{\textbf{\shortstack{Prediction\\Horizon}}} 
& \multicolumn{2}{c|}{\textbf{VRU Crosswalks}} 
& \multicolumn{2}{c|}{\textbf{Roundabout Entries}} 
\\
& \textbf{Precision} & \textbf{Recall} 
& \textbf{Precision} & \textbf{Recall} 
\\

\hline
\multirow{2}{*}{\textbf{1}} 
& 0.97 & 0.99      
& 0.98 & 0.98 \\
& 0.98 & 0.98 
& 0.98 & 0.98 \\
\arrayrulecolor{lightgray}\hline

\multirow{2}{*}{\textbf{2}} 
& 0.95 & 0.94      
& 0.96 & 0.93 \\
& 0.95 & 0.93 
& 0.96 & 0.92 \\
\arrayrulecolor{lightgray}\hline

\multirow{2}{*}{\textbf{3}} 
& 0.91 & 0.88    
& 0.85 & 0.80 \\
& 0.90 & 0.86 
& 0.88 & 0.86 \\
\arrayrulecolor{lightgray}\hline

\multirow{2}{*}{\textbf{4}} 
& 0.86 & 0.83      
& 0.67 & 0.63 \\
& 0.84 & 0.78 
& 0.81 & 0.79 \\
\arrayrulecolor{lightgray}\hline

\multirow{2}{*}{\textbf{5}} 
& 0.80 & 0.71     
& 0.56 & 0.47  \\
& 0.76 & 0.64
& 0.75 & 0.75  \\
\arrayrulecolor{black}\hline
\end{tabular}
\label{tab:occupancy}
\end{table}

With respect to occupancy, the dataset shows an imbalance concerning the occupation of entry and crosswalk zones: crosswalks are positive, i.e., occupied, in only 1.3\% of cases, reflecting the low amount of VRUs in the \textit{openDD} dataset. Entries are positive in 11.6\% of samples, as many vehicles leave the roundabout at the first opportunity. Precision is therefore essential to assess how reliably the model identifies true occupancies without causing false positives (i.e., unnecessary speed advisories by ROSA), while recall measures the model’s ability to detect all actual positives, minimizing missed occupancies (i.e., optimization opportunities).
Occupancy evaluation of model~(2) and model~(3) confirms high short-term accuracy for both crosswalk and entry zones (see Table \ref{tab:occupancy}).
At a one-second prediction horizon, both achieve near-perfect scores ($Precision \geq0.97, Recall \geq0.98$). As the prediction horizon increases, performance gradually decreases but remains robust up to three seconds ($Precision \geq0.85, Recall \geq0.80$).
Beyond that, uncertainty increases, with model~(2) showing a stronger decline in predictive performance for roundabout entries ($Precision=0.56, Recall=0.47$ at five seconds) compared to model~(3) ($Precision=Recall=0.75$), highlighting the benefit of exit intention modeling. For crosswalks, both models maintain predictive accuracy over five seconds, despite a gradual decrease reflecting the behavioral variability of VRUs.
Accuracy and F1 score remain high throughout all prediction horizons, though this may be inflated by natural imbalance in the real-world dataset with regard to the occupancy.
Overall, both models reveal predictive value over the full five-second horizon, enabling ROSA to provide proactive speed advisories based on inferred occupancy.

\section{Roundabout Optimized Speed Advisory}
We introduce the speed advisory in ROSA, designed to dynamically optimize
the speed of approaching vehicles based on real-time predictions of crosswalk and entry occupancy. It is based on physical principles, similar to GLOSA -- Green Light Optimized Speed Advisory -- which recommends an optimal speed based on upcoming signal changes to efficiently pass a signalized intersection at green phase (for more details see e.g., \cite{SchlampGLOSA23}). 
We assume a central unit collects past and current trajectory data from infrastructure or cooperative perception via FCOs. Prediction over $m = 5$ time steps can be either centralized, with occupancy information transmitted to ROSA-equipped vehicles, or decentralized, with trajectory data sent to the vehicles via V2X.

\subsection{Algorithm Description}
ROSA processes the following input parameters:
\begin{itemize}
\item[1.] The current vehicle speed $v$.
\item[2.] The distance to the VRU crosswalk $d_c$.
\item[3.] The distance to the roundabout's entry $d_{e}$.
\item[4.] The past and current traffic states, including kinematic and semantic data per agent.
\item[5.] The maximum prediction horizon $m$.
\end{itemize}
\medskip

As outlined in Algorithm~\ref{algo:ROSA}, ROSA computes the time-to-arrival $t_c$ at the VRU crosswalk based on current speed and distance. If $t_c$ falls within the prediction horizon~$m$, the traffic state is autoregressively predicted over all $m$ steps (see Section~\ref{sec:prediction}), allowing crosswalk occupancy to be assessed upon arrival. If clear, the vehicle is advised to maintain its current speed. If being occupied, a reduced optimal speed~$v_{opt}^{c}$ is calculated. 
Next, the time-to-arrival $t_{e}$ at the entry is determined based on the optimal crosswalk speed $v_{opt}^{c}$ and the remaining distance. 
If $t_{e} \leq m$, $v_{opt}^{c}$ is re-evaluated and, if needed, adjusted based on predicted entry occupancy. This two-stage optimization ensures the vehicle reaches both the crosswalk and entry when they are predicted to be clear, minimizing conflicts and improving efficiency when approaching and entering the roundabout.

Optimal speeds are calculated using basic kinematic equations~(\ref{Eq1}). Since occupancy is predicted only up to $m$ seconds ahead, and occupancy duration beyond $m$ is unknown, the algorithm targets arrival at $t+1$ ($t \in {t_c, t_e}$) to maximize the chance of conflict-free passage. Speed adjustments respect a maximum deceleration of 2 m/s², which may prevent immediate compliance with the advisory. ROSA operates at one-second intervals to address both vehicle dynamics and higher occupancy durations, continuously updating the optimal speed based on the current speed and predicted occupancy.

\begin{equation} \label{Eq1}  
    v_{opt} \enspace {=} \enspace \frac{2 \ast d}{t} - v
    \bigskip
\end{equation}

\begin{algorithm}
\caption{ROSA Algorithm} \label{algo:ROSA}
\begin{algorithmic}
\State Calculate time-to-arrival at crosswalk $t_{c}$ based on distance~$d_c$ and current speed $v$
\If{$t_c \leq$ prediction horizon $m$} 
    \State Autoregressively predict agent positions over $m$ steps
    \State Infer crosswalk and entry occupancy 
    \State Check crosswalk occupancy at $t_{c}$
    \If{CLEAR}
        \State Continue trip with current speed $v_{opt}^{c} = v$
    \Else
        \State Calculate optimal speed $v_{opt}^{c}$ for $t_{c} + 1$
    \EndIf
    \State \parbox[t]{\dimexpr\linewidth-\algorithmicindent}{%
    Calculate time-to-arrival at entry $t_{e}$ based on distance~$d_e$ and current speed $v$
    }
    \If{$t_{e} \leq m$}
        \State Check entry occupancy at $t_{e}$
        \If{CLEAR}
            \State Continue trip with crosswalk speed $v_{opt}^{e} = v_{opt}^{c}$
        \Else
            \State Calculate optimal speed $v_{opt}^{e}$ for $t_{e} + 1$
        \EndIf
    \EndIf
\EndIf
\end{algorithmic}
\end{algorithm}

\subsection{Evaluation Setup}
The effectiveness of ROSA is evaluated through a microscopic traffic simulation, modeling a roundabout environment using the Simulation of Urban MObility (SUMO)~\cite{behrisch2011sumo}. Real-world trajectories from the test dataset described in Section~\ref{dataset} are used to model realistic occupancy states, producing 6600 real-world scenarios. The evaluation focuses on a single ego vehicle approaching the roundabout from a fixed distance of 250~meters along a predefined route. The vehicle is assumed to be connected and fully automated, directly executing speed advisories. Nevertheless, the method is not limited to a single AV, but also generalizes to connected vehicles with human drivers and scales to all motorized road users in the scene.
As the ego vehicle approaches the roundabout, it is initially controlled by the default vehicle dynamics model implemented in SUMO. ROSA receives the relevant input from the simulation at one-second intervals and returns a speed advisory once the time-to-arrival conditions are met. The SUMO-controlled vehicle starts decelerating from 50 km/h around 100 meters before the entry, reflecting typical approaching behavior. Based on this deceleration, the time-to-arrival conditions are met and ROSA is triggered around 46~meters before the VRU crosswalk and 54~meters before the entry. With the two conflict zones about 8 meters apart, at least the first two ROSA executions optimize speed with respect to crosswalk occupancy only. Entry occupancy is included once its time-to-arrival falls within the prediction horizon. After passing both, control returns to SUMO.

By comparing trips with and without speed advisories based on predicted occupancy from model~(2) and model~(3), potential efficiency and safety improvements under real-world conditions are quantified.
Vehicle efficiency is evaluated for the Internal Combustion Engine (ICE\footnote{The EURO4 vehicle model, standardized in SUMO, is utilized.}) vehicle using performance metrics, including fuel consumption, CO$_2$ emissions, travel time, waiting time, and the number of stops -- averaged across all evaluation scenarios. Additionally, we evaluate a Battery Electric Vehicle (BEV) based on the same trajectories, where only energy consumption and emissions differ due to the vehicle type.
Safety is indirectly assessed through the reduction of vehicle stops, quantifying how ROSA minimizes vehicle-VRU and vehicle-vehicle conflicts. Especially in vehicle-VRU interactions, perceived safety from a VRU perspective improves with greater distances and reduced approach speeds.

In the test dataset, approximately 16\% of the scenarios exhibit potential for optimization, as the ego vehicle encounters an occupied crosswalk and/or roundabout entry. Based on this distinction, results are categorized into scenarios with and without optimization potential. Outcomes are reported separately for each category and the full test dataset. The evaluation of ROSA includes both our proposed prediction models from Section \ref{sec:prediction} and perfect foresight (ground truth) conditions to quantify the performance gap caused by prediction uncertainty.

\subsection{Results}

\renewcommand{\arraystretch}{1.2}
\begin{table*}[ht]
\centering
\captionsetup{justification=centering}
\caption{Efficiency and safety improvements of ROSA in 6600 real-world scenarios, averaged across optimizable, non-optimizable, and all scenarios. Results under prediction uncertainty (model (2) in the top row, model (3) in the bottom row) are shown on the left, with perfect foresight presented on the right.}
\begin{tabularx}{\textwidth}{|c|
>{\centering\arraybackslash}p{0.081\textwidth} |
>{\centering\arraybackslash}p{0.127\textwidth} |
>{\centering\arraybackslash}p{0.081\textwidth} |
>{\centering\arraybackslash}p{0.081\textwidth} |
>{\centering\arraybackslash}p{0.127\textwidth} |
>{\centering\arraybackslash}p{0.081\textwidth} |}
\hline
\multirow{3}{*}{\textbf{Performance Metric}} 
& \multicolumn{3}{c|}{\textbf{ROSA with Prediction Uncertainty}} 
& \multicolumn{3}{c|}{\textbf{ROSA with Perfect Foresight}} \\
& \makecell{Optimizable\\Scenarios} 
& \makecell{Non-Optimizable\\Scenarios} 
& \makecell{All\\Scenarios} 
& \makecell{Optimizable\\Scenarios} 
& \makecell{Non-Optimizable\\Scenarios} 
& \makecell{All\\Scenarios} \\
\hline      
\multirow{2}{*}{\textbf{BEV Energy Consumption}} 
& -9.24\% & +0.31\% & -1.21\%
& \multirow{2}{*}{-16.56\%} & \multirow{2}{*}{+/-0.00\%} & \multirow{2}{*}{-2.63\%} \\
& -10.16\% & +0.10\% & -1.53\% & & & \\
\arrayrulecolor{lightgray}
\hline
\multirow{2}{*}{\textbf{Fuel Consumption \& CO$_2$ Emissions}} 
& -4.14\% & +1.23\% & +0.38\%
& \multirow{2}{*}{-7.75\%} & \multirow{2}{*}{+/-0.00\%} & \multirow{2}{*}{-1.22\%} \rule{0pt}{2.5ex} \\
& -4.71\% & +0.63\% & -0.22\% & & & \\
\hline
\multirow{2}{*}{\textbf{Travel Time}} 
& -2.94\% & +0.43\% & -0.1\%
& \multirow{2}{*}{-5.28\%} & \multirow{2}{*}{+/-0.00\%} & \multirow{2}{*}{-0.84\%} \rule{0pt}{2.5ex} \\
& -3.34\% & +0.23\% & -0.34\% & & & \\
\hline
\multirow{2}{*}{\textbf{Waiting Time}} 
& -61.22\% & +1.26\% & -8.66\%
& \multirow{2}{*}{-94.91\%} & \multirow{2}{*}{+/-0.00\%} & \multirow{2}{*}{-15.07\%} \rule{0pt}{2.5ex} \\
& -66.30\% & +0.83\% & -9.83\% & & & \\
\hline
\multirow{2}{*}{\textbf{Number of Stops}} 
& -57.87\% & +1.26\% & -8.13\%
& \multirow{2}{*}{-93.37\%} & \multirow{2}{*}{+/-0.00\%} & \multirow{2}{*}{-14.83\%} \rule{0pt}{2.5ex} \\
& -63.31\% & +0.83\% & -9.36\%  & & & \\
\arrayrulecolor{black}
\hline
\end{tabularx}
\label{table:ROSA}
\end{table*}

The evaluation demonstrates ROSA’s full potential under perfect foresight (see Table \ref{table:ROSA}, right). In the 16\% of scenarios requiring optimization, BEV energy consumption is reduced by around 17\%, fuel consumption and CO$_2$ emissions by 8\%, travel time by 5\%, waiting time by 95\%, and stops by 93\%. The marked reduction in stops also indicates a minimization of conflicts between vehicles and VRUs. Averaged across all 6600 scenarios, this yields reductions of around 1–3\% in energy and emissions, 0.8\% in travel time, and 15\% in both waiting time and stops -- demonstrating that even infrequent optimizations lead to measurable system-wide improvements.

With predicted trajectories instead of ground truth, ROSA still achieves substantial gains in optimizable scenarios, with model~(3) (motion dynamics and exit intention) providing only a slight performance edge over model~(2) (motion dynamics only).
As shown in Table \ref{table:ROSA} (left), energy consumption is reduced by around 10\%, emissions by 5\%, travel time by 3\%, waiting time by 66\%, and stops by 63\%. Compared to perfect foresight, this reflects a performance drop of about one-third across all key metrics, mainly due to missed optimization opportunities from false negatives in occupancy prediction, i.e., undetected conflicts with vehicles or VRUs.
In non-optimizable scenarios, prediction uncertainty can cause false positives, i.e., falsely predicted occupancy. These cases occur more often with model~(2), leading to greater performance losses (energy +0.3\%, emissions +1\%, travel time +0.4\%, waiting time and stops +1\%) compared to model~(3), where increases remain below 0.8\%.
To conclude, false positive losses are negligible, with the main performance decline resulting from false negatives. Nevertheless, both prediction models still enable ROSA to enhance safety and efficiency in diverse real-world traffic scenarios, with model~(3) providing slightly better outcomes compared to model~(2).

\section{Discussion and Conclusion}
We present ROSA -- Roundabout Optimized Speed Advisory -- a system for coordinated speed guidance in multimodal and mixed traffic at roundabouts. Designed to support both automated and human-driven vehicles, ROSA ensures safe and efficient interaction with both conflicting VRUs and vehicles. At its core, ROSA integrates a Transformer-based, multi-agent trajectory prediction model that captures inter-agent dependencies, addressing the challenge of joint prediction for VRUs and vehicles in dynamic, interaction-dependent roundabout environments. 
Based on predicted conflicts with VRUs and vehicles, ROSA provides real-time, proactive speed advisories for vehicles approaching and entering roundabouts. 

Results show that our prediction model incorporating motion dynamics significantly outperforms previous methods (ADE: 1.29m, FDE: 2.99m at a 5s prediction horizon). Integrating route intention yields additional improvements (ADE:~1.10m, FDE: 2.36m), demonstrating the value of connected vehicle data.
Translating predicted trajectories into crosswalk and entry occupancy, both models maintain accuracy over the full five-second horizon, enabling proactive speed advisories through ROSA.
ROSA successfully improves safety and vehicle efficiency, even under prediction uncertainty, with BEVs benefiting more than conventional ICE vehicles. Performance losses mainly stem from false negatives in occupancy prediction, preventing necessary optimizations, while false positives are rare and have minimal impact. The results confirm ROSA’s practical value, with even greater benefits expected in higher traffic volumes.

To conclude, this work addresses the challenges of automated driving in multimodal traffic at roundabouts. Through coordinated speed guidance for vehicles in VRU-prioritized settings, ROSA effectively resolves conflicts, reduces delays and emissions, and improves roundabout capacity. Its proactive speed reduction also enhances perceived safety for VRUs, making roundabouts more attractive for non-motorized users. ROSA therefore balances the trade-off between safety and efficiency, paving the way for sustainable and trustworthy mobility in diverse traffic ecosystems. Designed for mixed traffic, it delivers immediate benefits without requiring full fleet penetration. Its data-driven architecture without prior assumptions scales to all motorized road users and generalizes to diverse settings, such as roundabouts with varying layouts and traffic rules, and even unsignalized intersections. However, stable performance in broader contexts requires more trajectory prediction datasets with sufficient VRU representation.

Future research might expand ROSA to optimize the speed of multiple vehicles in a cooperative way, where finding the optimal solution within a high-dimensional decision space is crucial, e.g., with Reinforcement Learning. The focus of this work was to analyze ROSA's potential based on real-world trajectories. In further research, this could be complemented by an evaluation of varying traffic conditions, either through additional real-world datasets or simulations. Moreover, ROSA~-- as a combination of trajectory prediction and traffic coordination at roundabouts -- can be used as an example function to evaluate the role of different perception approaches (e.g., infrastructure compared to FCOs), V2X related uncertainty and latency, and human compliance for coordinating multimodal and mixed traffic. Real-world experiments could validate our findings and help transition to practical deployment.
\newline

\end{document}